\documentclass[12pt]{article}
\usepackage{cite}
\usepackage{graphicx}
\usepackage{amssymb,amsmath,amsfonts,amsthm}
\usepackage{epstopdf}
\usepackage[normalem]{ulem}
\usepackage{comment}
\usepackage{hyperref}
\title{\Large \bf{Analysis of the $e^{+}e^{-}\rightarrow\pi^{0}\gamma$ process using anomaly sum rules approach}}
\author{
S.~P.~Khlebtsov,$^{1}$
A.~G.~Oganesian,$^{1,2}$
O.~V.~Teryaev,$^2$
\footnote{
  Electronic addresses:
  \href{mailto:khlebtsov@itep.ru,}{khlebtsov@itep.ru},
  \href{mailto:armen@itep.ru,}{armen@itep.ru},
  \href{mailto:teryaev@theor.jinr.ru}{teryaev@theor.jinr.ru}.
}
 \vspace{12pt} \\
\it \small $^1$Institute of Theoretical and Experimental Physics, 117218,  Moscow, Russia\\
\it \small $^2$ Bogoliubov Laboratory of Theoretical Physics,\\
\it \small Joint Institute for Nuclear Research, 141980, Dubna, Russia\\
       }
\date{}
\begin{document}
\maketitle

\begin{abstract}
The process $e^{+}e^{-}\rightarrow\gamma^{*}\rightarrow
\pi^{0}\gamma$ was considered using time-like pion transition form
factor, obtained in the approach of the Anomaly Sum Rules(ASR). The
total cross section and angular distribution of the process was
calculated. As the result of the comparison with the data it was
shown that ASR approach provides their good description in the
regions far from the pole. Also there was proposed a method allowing
to give reasonable description of data in the region of pole within
the ASR approach. The strong restrictions for the parameters of the
modified ASR approach were obtained.
\end{abstract}

\newpage
\section{Introduction }

The transition form factor(TFF) of the pion is attracting the great
interest last years. In particular, it is related to the pion radius
at small virtualities of the photon.
%so that one  one can obtain radius of a particle, knowing it form factor.
It is used for description of the hadrons-photon interactions, and
within it, it is possible to match predictions of the pQCD and
non-perturbation methods. 

The space-like region of the transition form factor of the $\pi^0$
is quite good investigated. The available experimental data cover a
fairly wide range of $Q^2$. The CELLO\cite{cello} and
CLEO\cite{cleo} collaborations measured it in the intervals $0.7-2.2
\ GeV^2$ and $1.6-8.0 \ GeV^2$ , respectively. The BaBar\cite{babar}
and Belle\cite{belle} move the range of $Q^2$ up  to 40 $GeV^2$. At
the $1.0\leq Q^2\leq9$ all the experiments show same behaviour and
agree with theoretical expectations\cite{lepage}. From the other side, there is well known
contradiction between BaBar\cite{babar} and Belle\cite{belle}
results of the measurement of the $\pi^0$ TFF at the region of the
large space-like photon virtualities. At the same time,
in the region of low $Q^2$ and in the time-like region the number of
direct measurements of the pion TFF is quite small. The more precise
data are expected in the future from BES-III\cite{bes3} and
KLOE-2\cite{kloe2} collaborations.

Our paper is dedicated to the pion transition form factor in the
time-like region. In this region pion TFF can be studied in the
process $e^{+}e^{-}\rightarrow\gamma^{*}\rightarrow \pi^{0}\gamma$.
The SND\cite{datasnd} (the most accurate data on the process
$e^{+}e^{-}\rightarrow\pi^{0}\gamma$ to date) and CMD2\cite{datacmd}
experiments had collected the data that cover the range $0.6-1.38 \
GeV$ of $\sqrt{s}$. Theoretically, the process was studied in the
work \cite{HKLNS}, with using the methods of the dispersive theory.
In the future, CLAS\cite{clas} will provide more precise data of the
direct measurements of the time-like pion TFF.

Recently a new method was offered  to describe the pion form factor
based on the dispersive representation of the axial anomaly
\cite{hor-ter},\cite{kot2011} (see also review \cite{IOFFE2}). This method has an advantage of being
model-independent and not relying on the QCD factorization. Also in
this approach the pion TFF can be described in the whole region of
the photon virtualities $Q^2=-q^2$. In the paper \cite{kot2011} the
pion TFF was obtained in the whole space-like region $Q^2>0$. Later
it was analytically continued to the time-like region
\cite{KOT2013}. As soon as this method is based on the exact
relation, implied by the axial anomaly it provides very powerful
tool to study pion (and other pseudoscalar mesons) TFFs both in
space-like and time-like regions.

The goal of this work is to check predictions of the ASR method in
the time-like region at low $q^2$ using the SND and CMD2
experimental data.

The paper is organized as follows: the section \ref{tcsc} contains a
brief description of the main steps of the pion TFF calculation and
the calculation of total cross section and angular distribution of
the process $e^{+}e^{-}\rightarrow\gamma^{*}\rightarrow
\pi^{0}\gamma$; in the next section \ref{comp} the obtained
expressions are compared with the experimental data: the subsection
\ref{general} comprises analysis of the situation far from the
poles, while the subsections \ref{peak1} and \ref{peak2} are
dedicated to the attempts to find modifications to the expression of
the pion TFF in order to describe data peaks, within the ASR
approach; in the last section \ref{last} we summarize the obtained
results.

\section{Total cross section calculation} \label{tcsc}

\begin{figure}[h!]
\centerline{
\includegraphics[width=0.6\textwidth]{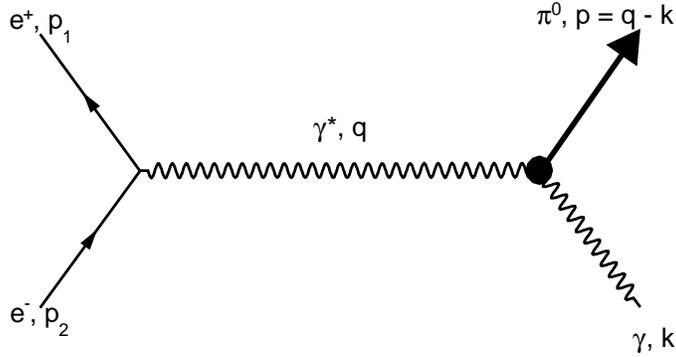}}
\caption{Feynman diagram of the process.} \label{fig:1}
\end{figure}

Let us consider a process
$e^{+}e^{-}\rightarrow\gamma^{*}\rightarrow\pi^{0}\gamma$. It can be
expressed in terms of the pion TFF $F(q^2)$, which is defined as:
\begin{equation}\label{TFF}
 \int d^{4}x e^{ikx} \langle \pi^0(p)|T\{J_\mu (x) J_\nu(0)
\}|0\rangle = \epsilon_{\mu\nu\rho\sigma}k^\rho q^\sigma F(q^2) \;,
\end{equation}
where $k,q$ are momenta of photons, $p=q-k$, and two electromagnetic
currents $J_{\mu}=\sum\limits_{i=u,d,s} e_i \bar{q_i}\gamma_\mu
q_i$. In what following all the expressions will be written in the unit of electron charge $|e|=1$, while in the equations for cross sections and amplitudes explicit dependence of electron charge is restored. One photon is real $(k^2=0)$ and the other one is virtual. The
Feynman diagram is shown on the Fig.\ref{fig:1}.

The expression for the pion TFF was obtained in approach based on
the dispersive representation of the axial anomaly in the works
\cite{kot2011},\cite{KOT2013}. Here we briefly recall the main steps
of the method.

The vector-vector-axial triangle graph amplitude, where the axial
anomaly occurs, contains an axial current $J_{\alpha 5}$ and two
electromagnetic currents $J_{\mu}=\sum\limits_{i=u,d,s} e_i
\bar{q_i}\gamma_\mu q_i$,
\begin{equation} \label{VVA}
T_{\alpha \mu\nu}(k,q)=\int d^4 x d^4 y e^{(ikx+iqy)} \langle 0|T\{
J_{\alpha 5}(0) J_\mu (x) J_\nu(y) \}|0\rangle,
\end{equation}
where $k$ and $q$ are the photons momenta. In what follows, we limit
ourselves to the case when one of the photons is on-shell ($k^2=0$).
As it was shown in the paper \cite{hor-ter} the imaginary part $A_3$
of the of the invariant amplitude at the tensor structure $k_{\nu}
\varepsilon_{\alpha \mu \rho \sigma}k^{\rho} q^{\sigma}$ in the
variable $ (k+q)^2= s > 0$ satisfy the following relation:
\begin{equation}\label{asr}
\int_{0}^{\infty} A_{3}(s,q^{2}; m_i^{2}) ds =
\frac{1}{2\pi}\frac{1}{\sqrt{2}}.
\end{equation}
The relation \eqref{asr} is exact: $\alpha_s$ corrections are zero
and it is expected that all nonperturbative corrections are absent
as well (due to 't Hooft's principle \cite{hor-ter,tHooft}). Note,
in the original paper \cite{hor-ter} the eq.\eqref{asr} was
obtained for the space-like photon $(q^2<0)$. Later in the paper
\cite{KOT2013} analytical continuation to the time-like region was
developed.

Supposing that $A_3$ decreases fast enough at $\lvert q^2 \rvert \to
\infty$ and is analytical everywhere except the cut
$q^2\in(0,+\infty)$, it was found

%\begin{equation} \label{a3disp}
%A_3^{(a)}(s,q^2)=\frac{1}{2\pi }\int_{0}^{\infty}dy\frac{\rho^{(a)}(s,y)}{y-q^2+i\epsilon},
%\end{equation}
%where  $\rho^{(a)}=2Im_{q^2} A_3^{(a)}$. Then, the ASR (\ref{asr}) for time-like $q^2$ is given by the double dispersive integral:

%\begin{align} \label{asr-1}
%\int_{0}^{\infty}ds\int_{0}^{\infty}dy  \frac{\rho^{(a)}(s,y)}{y-q^2+i\epsilon}=N_c C^{(a)}, \; a=3,8.
%\end{align}
%Note, that generally speaking, the order of integration cannot be interchanged.

\begin{align} \label{asr-re}
p.v.\int_{0}^{\infty}ds\int_{0}^{\infty}dy
\frac{\rho(s,y)}{y-q^2}=\frac{1}{\sqrt{2}},\\ \label{asr-im}
\int_{0}^{\infty}ds \rho(s,q^2)=0,\; .
\end{align}
where  $\rho=2Im_{q^2} A_3$. Saturating the lhs of the three-point
correlation function \eqref{VVA} with the resonances in the axial
channel, singling out the first (pion) contribution and replacing
the higher resonance's contributions with the integral of the
spectral density, the ASR in the time-like region \eqref{asr-re}
leads to
\begin{equation} \label{qhd3}
\pi f_{\pi}Re F_{\pi\gamma}(q^2)+ \int_{s_3}^{\infty} A_{3}(s,q^{2})
ds  =\frac{1}{2\pi}\frac{1}{\sqrt{2}},
\end{equation}
where $s_3$ is duality region of the pion in the isovector channel,
the definition of the TFFs $F_{\pi\gamma}$ is \eqref{TFF}, and the
meson decay constant $f_{\pi}$ is,
\begin{align} \label{def_f}
\langle& 0|J_{\alpha 5}(0) |\pi^0(p)\rangle= i p_\alpha f_{\pi},
\end{align}
where $J_{\alpha
5}(0)=\frac{1}{\sqrt{2}}(\bar{u}\gamma_{\alpha}\gamma_5u-\bar{d}\gamma_{\alpha}\gamma_5d)$.
As the integral of $A_3$ in eq.\eqref{qhd3} is over the region
$s>s_3$, we expect that nonperturbative corrections to $A_3$ in this
region are small enough and we can use the one-loop expression for
it.

Then the ASR leads to the pion TFF:
%\begin{align}
%Re F_{\pi\gamma}(q^2)=&\frac{N_c C^{(3)}}{2\pi^2f_\pi}\left[p.v.\int_{0}^{s_3}ds \int_{0}^{\infty} dy \frac{\rho^{(a)}(s,y)}{y-q^2} \right]= \nonumber \\ \label{f3m-2}
%&\frac{1}{2\sqrt{2}\pi^2f_{\pi}}\frac{s_3}{s_3-q^2}.
%\end{align}
%Thus,
\begin{equation}
F(q^2)=\frac{1}{2\sqrt{2}\pi^2f_{\pi}}\frac{s_3}{s_3-q^2}.
\label{ff}
\end{equation}
As were discussed in the papers \cite{kot2011},\cite{KOT2013}, this
result is valid in both time-like and space-like regions (expect the
pole $q^2=s_3$).

The numerical value of $s_3$ was obtained in the limit
$-q^2\rightarrow\infty$ of the space-like ASR \cite{kot2011}, $s_3 =
4\pi^2f^2_{\pi}=0.67 \ GeV^2 \pm 10\% $. This expression coincides
with the one obtained earlier from the two-point correlator
analysis\cite{rad} and is close to the numerical value obtained from
two-point sum rules\cite{SVZ}. In the recent analysis light cone and
anomaly sum rules predictions were compared \cite{OPST}, and it was
shown that $s_3$ is indeed approximately constant with the accuracy
about $\pm10\%$. At same time it was noted that in the region of
small $q^2$ the value $s_3=0.61 \ GeV^2$ is more preferable.

As we can see from \eqref{ff} the pion time-like TFF has a pole at
$q^2=s_3$, which is numerically close to $m^2_{\rho}\approx0.6 \ GeV^2$
and to $m_\omega^2\approx0.61 \ GeV^2$. The pole behavior (which
corresponds to zero width of the ρ meson) appeared since we used
the one-loop approximation for $A_3$, neglecting the possible
dependence of $s$ on $Q^2$ and final-state interactions. Therefore,
the eq. (\ref{ff}) can be used not too close to the pole $q^2 =
s_3$. The effect of the finite width can be estimated if one takes
into account small corrections: as perturbative($\alpha_s^2$ and
higher) as non-perturbative; and also the small effects of mixing.

One can write down the amplitude $M_{\pi\gamma}$ as:
\begin{equation}
M_{\pi\gamma}  =
ie^3\bar{u}\gamma^{\alpha}u\frac{g_{\alpha}^{\mu}}{q^2}\epsilon_{\mu\nu\rho\sigma}k^{\rho}q^{\sigma}F(q^2)(e^{\nu})^{*}
\label{amp}
\end{equation}
Neglecting masses and summing over polarisations, the square of the
amplitude takes form:
\begin{equation}
|M_{\pi\gamma}|^2 =
\frac{e^6}{4}Sp[\not{p_{1}}\gamma^{\mu}\epsilon_{\mu\nu\rho\sigma}k^{\rho}q^{\sigma}\not{p_{2}}\gamma^{\mu'}\epsilon_{\mu'\nu'\rho'\sigma'}k^{\rho'}q^{\sigma'}g^{\nu\nu'}]\frac{|F^2(q^2)|}{q^4}.
\label{amp_sq}
\end{equation}
Total cross section has the form:
\begin{equation}
\sigma=\frac{2}{3}\pi^2\alpha_{QED}^3|F^2(q^2)|.
\label{asr_cs}
\end{equation}
Finally, substituting eq.\eqref{ff} to \eqref{asr_cs} we obtain expression for the total cross section:
\begin{equation}
\sigma_{theor}=\frac{\alpha_{QED}^3}{12\pi^{2}f_{\pi}^2}\frac{s_{3}^2}{(s_{3}-q^2)^2}.
\label{asr_cs_theor}
\end{equation}
And the expression for angular distribution has a rather common form
implied by angular momentum conservation:
\begin{equation}
\label{angular}
\frac{d\sigma}{d\cos\theta}\big|_{\pi}^0=\frac{\alpha_{QED}^3}{32\pi^2f_{\pi}^2}\frac{s_{3}^2}{(s_3-q^2)^2}(1+\cos^2\theta).
\end{equation}

\section{Comparison with data}\label{comp}
\subsection{Comparison with experiment out of poles}\label{general}
\begin{figure}[h!]
\centerline{
\includegraphics[width=1.0\textwidth]{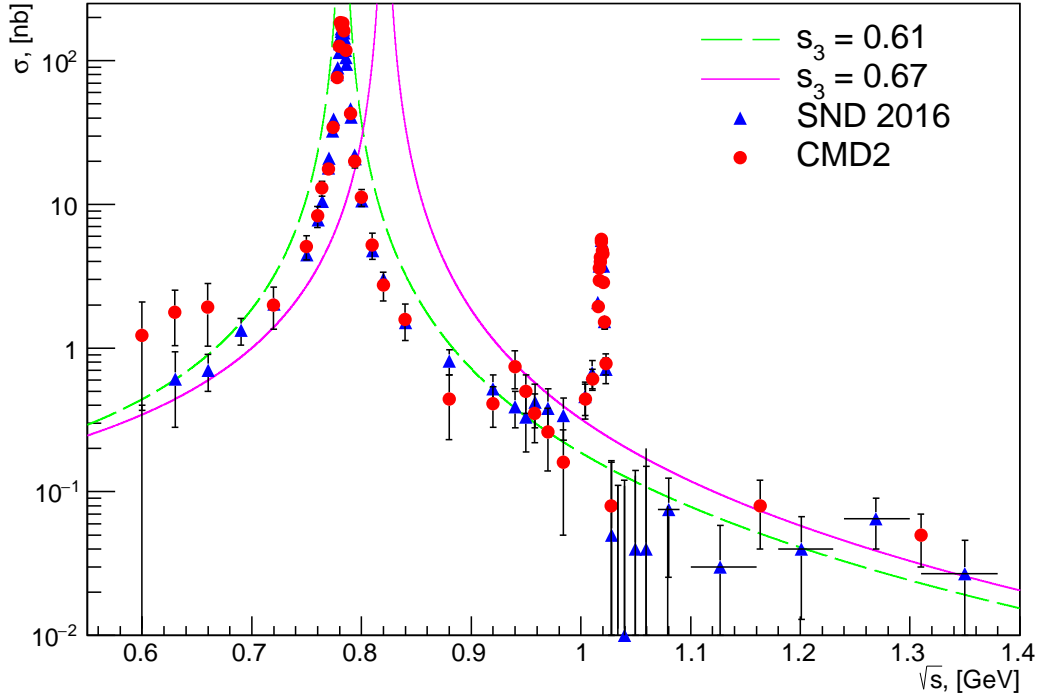}}
\caption{Total cross section $\sigma_{theor}$ vs. SND2016 and CMD2
\cite{datasnd,datacmd} data.} \label{fig:2}
\end{figure}

In  this section we will compare theoretical results of the ASR
approach \eqref{asr_cs_theor} with SND2016 and CMD2 experimental data
\cite{datasnd},\cite{datacmd}. The curves corresponding to
eq.\eqref{asr_cs_theor} with the lower and upper limits $s_3 = 0.61, 0.67
\ GeV^2$ are shown (by the dashed and solid lines) on the Fig.\ref{fig:2}.

Let us emphasize, that the eq.\eqref{ff} indicates the $\rho-\omega$
meson resonance( in the zero width approximation) position, but do
not indicate the existence of the second resonance, corresponding to
the $\phi$ meson mass. The reason of this is that the equation of
the pion TFF was obtained for the isovector channel of the axial
current and it does not take into account the possible effects of
mixing. So the hadrons including s-quarks can't be accounted in such
approximation. Discussion and possible treatment of this will be
done later in the next sections.

Note, that from the Fig.\ref{fig:2} one can see curve with $s_3=0.67
\ GeV^2$(solid line) describes data much worse, than with $s_3=0.61
\ GeV^2$(dashed line). This one coincides with the results of the
matching Light-Cone Sum Rules(LCSR) and ASR approaches\cite{OPST} in
the space-like region, where, as it was mentioned earlier, it was
shown that at the region of small $q^2$ the value $s_3=0.61 \ GeV^2$
is more preferable, while within the $\pm10\%$ error of the $s_3$
calculation the value $s_3 = 0.67 \ GeV^2$ also agrees with the
experiment. But in the time-like region of the $q^2$, as can be seen
from Fig.\ref{fig:2}, the equation for the pion TFF \eqref{ff}, and
the corresponding total cross section \eqref{asr_cs_theor}, is much more
sensitive to the value of the $s_3$, and the value $s_3 = 0.67 \ GeV^2$ has
much worse agreement with the experiment, than $s_3=0.61 \ GeV^2$.
Thus, the time-like region is a kind of a microscope for analysis of
the parameters of the axial channel in the space-like region. So we
can expect, that analysis of the pion TFF in the time-like region of
$q^2$ allows to clarify the values of the small corrections, which
one are hard to determine in the space-like region.
%Обратите внимание из рис 2 видно что с3=0.67 описыватье данные существенно хуже, чем при с-3 =0.61. Это согласуется с результатами работы матчинг, где было показаено, что при малых ку квадрат с-3 = 0ю61 предпочтительное, чем с-3 =0.67. Однако, в работе матчинг, где аналищз был проведени для данных в спэйлайке, величина с-3 = 0.67 в пределах десяти процентной ошибки тоже давала согласие с экспериментом. Сейчас мы видим, что в тайм-лайке Однако анализ в тамй-лайке оказывается намного более чувствительным к величиние с-3 и с-3 = 0.67 намного хуже описывает эксперимент чем величина 0.61. таким образом видно, что тайм лайк является своеобразным микроскопом для изучения параметров аксиального канала в спэйслайке. таким образом мы можем надеятся, что анализ в тайм-лайке позволит учтонит величины малых поправок, которые затруднительно определить в спэйс-лайке.

The fact that the pole of the eq.\eqref{asr_cs_theor} $q^2=s_3=0.61$ is
close to the $m_{\rho}^2\approx0.60 \ GeV^2, m_{\omega}^2\approx0.61 \ GeV^2$ is
quite interesting. It actually means that from the parameters of the
axial channel one can obtain spectrum of masses in the vector
channel. The tendency of $s_3$ variation in the space-like channel
may be attributed to the effect of the pole in the time-like
channel, requiring that
%;;
$$s(m_V^2)=m_V^2.$$
%Thus, we see that the theoretical curve has the pole at $\sqrt{s}\approx0.8 \ GeV$, which corresponds to the $\rho$ and $\omega$ meson masses.
Clearly, at present accuracy one cannot distinguish $\rho$ and
$\omega$ masses.

It seems that more accurate analysis of the anomaly sum rule
\eqref{asr-re}, which will takes into account the effects of
$\pi,\eta,\eta'$ mixings (as well as perturbative and
non-perturbative small corrections), can provide a better
description of the experimental data, in particular of the second
peak on the Fig. \ref{fig:2}. Theoretical estimations show, that if
one includes the effects of the mixing into the calculation of the
pion TFF, then one obtains more complicated expression than the
eq.\eqref{ff}. It should be a linear combination of terms of the
type of  the eq.\eqref{ff} and each of them will have its own value
of $s_3$. This work is now in progress.

In the next chapter we discuss a modification of the eq.\eqref{ff} in
order to describe the peaks and estimate how good this approximation
is be able to describe experimental data. Firstly, we consider the
first peak and further generalize the result to the second one.

\subsection{$\rho$-$\omega$ peak}\label{peak1}

Let us consider the first experimental peak, which one is
corresponds to $\rho$-$\omega$ resonance. Here we perform fits of
the data below $1.0 \ GeV.$ The equation of the pion
TFF \eqref{ff} was obtained using zero width of $\rho$-meson, so the
$ImF(q^2)\sim\delta(s_3-q^2)$. To get right description of the data,
one should add to the denominator of \eqref{ff} term, which should
gives resonance corresponding to the data peak. It means that the
$\rho$-meson should have finite width, so the $ImF(q^2)$ will not
have such a trivial form.

\begin{figure}[h!]
\includegraphics[width=1.0\textwidth]{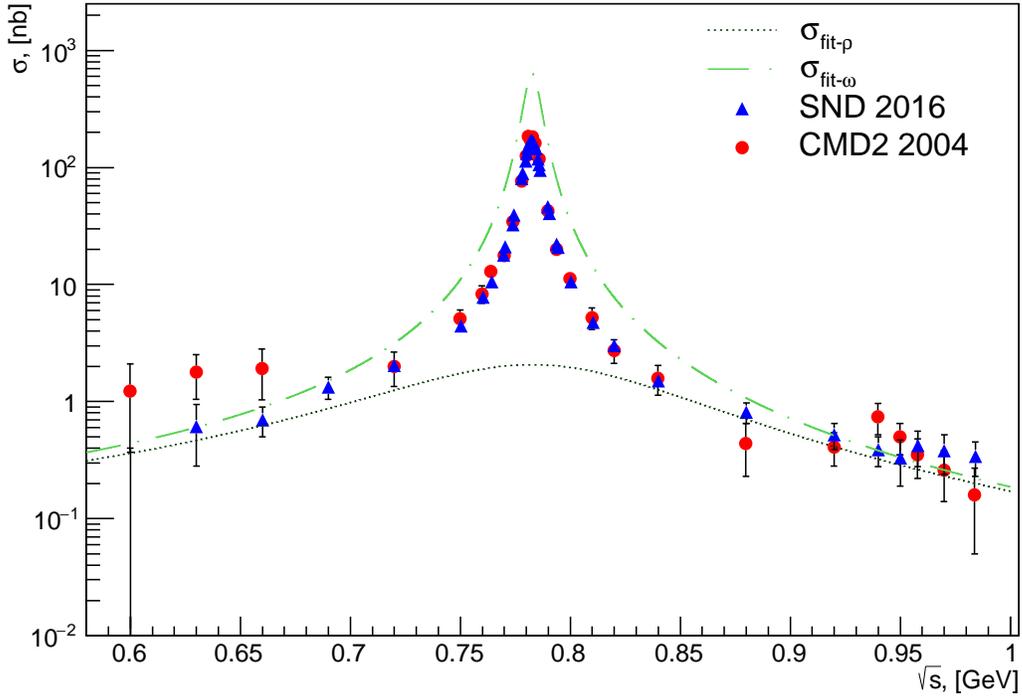}
\caption{Fits of the $\rho-\omega$ peak with 1 resonance.}
\label{fig:3}
\end{figure}

We modify \eqref{ff} by adding to denominator the term of the form
$im_v\Gamma_v$, so the equation for pion time-like TFF takes form
similar to the relativistic Breit-Wigner amplitude:
\begin{equation}
F(q^2)=\frac{1}{2\sqrt{2}\pi^2f_{\pi}}\frac{s_3}{s_3-q^2+im_v\Gamma_v}.
\label{mtff}
\end{equation}
Substituting to \eqref{amp_sq} the modified pion TFF equation
\eqref{mtff}, with the values of the $m_v, \Gamma_v, s_3=m_v^2$
corresponding to the $\rho$ and $\omega$ mesons( we take the
averaged one values of masses and widths of the $\rho$ , $\omega$
mesons from PDG \cite{PDG}: $m_{\rho}=0.77526 \ GeV,
\Gamma_{\rho}=0.149 \ GeV, m_{\omega}=0.78265 \ GeV,
\Gamma_{\omega}=0.00849 \ GeV.$), and doing simple calculations, we
obtain two fits for total cross sections:
\begin{equation}
\sigma_{fit-\rho,\omega}=\frac{\alpha_{QED}^3}{12\pi^{2}f_{\pi}^2}\frac{s_{3\rho,\omega}^2}{(s_{3\rho,\omega}-q^2)^2+m_{\rho,\omega}^2\Gamma_{\rho,\omega}^2}.
\label{fit_cs1}
\end{equation}
Result is shown on Fig.\ref{fig:3} for $\rho$ and $\omega$ cases
(dotted and dot-dashed lines, correspondingly). It is clearly seen
that both of this cases  have poor description of the experimental
data. The description is also not improved when the more complicated
models of the single resonance are applied.

Thus we can assume, that reasonable description of the first
experimental peak can be done by the to two resonance
parametrisation. If one takes linear combination of amplitudes type
of \eqref{mtff}, so, that the pion TFF will be a sum of the two
terms, where each one has the values of the $m_v, \Gamma_v,
s_{3v}=m_v^2$  corresponding to the $\rho$ and $\omega$ mesons:
\begin{equation}
\label{2resff}
F(q^2)=\frac{1}{2\sqrt{2}\pi^2f_{\pi}}\left(\alpha\frac{s_{3\rho}}{s_{3\rho}-q^2+im_{\rho}\Gamma_{\rho}}+\beta\frac{s_{3\omega}}{s_{3\omega}-q^2+im_{\omega}\Gamma_{\omega}}\right).
\end{equation}
Then total cross section takes form:
\begin{eqnarray}\label{2rescs}
\sigma_{fit-2resonances} & = & \frac{\alpha_{QED}^3}{12\pi^{2}f_{\pi}^2}\Big(\alpha^2\frac{s^2_{3\rho}}{(s_{3\rho}-q^2)^2+m_{\rho}^2\Gamma_{\rho}^2}+\beta^2\frac{s^2_{3\omega}}{(s_{3\omega}-q^2)^2+m_{\omega}^2\Gamma_{\omega}^2}+  {}\nonumber \\
             &   & +2\alpha\beta s_{3\rho}s_{3\omega}\frac{(s_{3\rho}-q^2)(s_{3\omega}-q^2)+m_{\rho}\Gamma_{\rho}m_{\omega}\Gamma_{\omega}}{((s_{3\rho}-q^2)^2+m_{\rho}^2\Gamma_{\rho}^2)((s_{3\omega}-q^2)^2+m_{\omega}^2\Gamma^2_{\omega})}\Big).
\end{eqnarray}
Note, due to that far the pole one should obtain formula \eqref{ff},
so $\alpha+\beta\approx1$.

\begin{figure}[h!]
\includegraphics[width=1.0\textwidth]{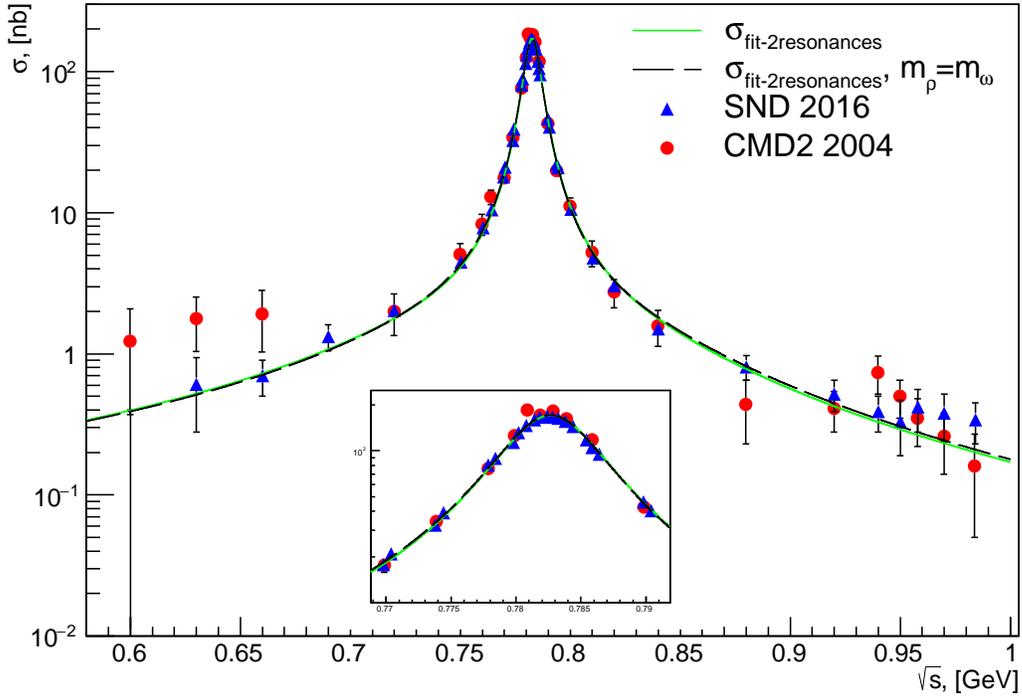}
\caption{Fits of the $\rho-\omega$ peak with 2 resonances. Insert is zoom in of the peak.} \label{fig:4}
\end{figure}

The result is shown on the Fig.\ref{fig:4}(solid line), the values of the fit parameters are: $\alpha = 0.52,\ \beta = 0.49;$ and the values of $\chi^2$ are in the Table\ref{tbl1}.

\begin{table}[h!]
\caption{Values of $\chi^2$ for fit with 2 resonances.}
\label{tbl1}
\begin{center}
\begin{tabular}{|c|c|c|c|}
\hline
 & \textbf{CMD2} & \textbf{SND2016} & \textbf{CMD2+SND2016} \\
\hline
$\chi^2/d.o.f.$ & 2.29 & 1.46 & 1.71 \\
\hline
\end{tabular}
\end{center}
\end{table}

From other side, such small value as the difference between $\rho$ and $\omega$ mesons masses is clearly out of the the accuracy of the ASR approach. That's why we perform one more fit using \eqref{2resff}, supposing  that $m_{\rho} = m_{\omega}$ and $s_{3\rho}=s_{3\omega}=m_{\rho}^2=m_{\omega}^2\approx0.61 \ GeV^2$, but with PDG \cite{PDG} values for widths: $\Gamma_{\rho}=0.149 \ GeV, \Gamma_{\omega}=0.00849 \ GeV.$ The result is shown on the Fig.\ref{fig:4}(dashed line), the values of the fit parameters are the same: $\alpha = 0.52,\ \beta = 0.49;$ and the values of $\chi^2$ are in the Table\ref{tbl2}.
%and the corresponding total cross section \eqref{2rescs}, with $m_{\rho} = m_{\omega}=0.78265 \ GeV$ and $s_{3\rho}=s_{3\omega}=m_{\omega}^2\approx0.61 \ GeV^2$,
%within the accuracy of the ASR approach one cannot distinguish $\rho$ and $\omega$ mesons masses,
%Due to that the equation of pion TFF \eqref{ff} was obtained on condition that $m^2_{\rho}=m^2_{\omega}=s_3$, so in our approach we can't distinguish $\rho$ and $\omega$ mesons. }

\begin{table}[h!]
\caption{Values of $\chi^2$ for fit with 2 resonances with $m_{\rho} = m_{\omega}$.}
\label{tbl2}
\begin{center}
\begin{tabular}{|c|c|c|c|}
\hline
 & \textbf{CMD2} & \textbf{SND2016} & \textbf{CMD2+SND2016} \\
\hline
$\chi^2/d.o.f.$ & 2.12 & 1.07 & 1.42 \\
\hline
\end{tabular}
\end{center}
\end{table}

As one can see from the Fig.\ref{fig:4} and Tables \ref{tbl1}, \ref{tbl2}, the both variants have very good agreement with data. Note, that the second one describes data even better than the first one.

In the limit of real photon $q^2\rightarrow0$ and neglecting $\Gamma_{\rho},\Gamma_{\omega}$ the pion TFF takes form:
$$F(0)=\frac{1}{2\sqrt{2}\pi^2f_{\pi}}(\alpha+\beta).$$%(in units electron charge $|e|=1$)
And correspondingly one can find 
$$\Gamma(\pi^0\rightarrow2\gamma)\approx7.9 \ eV,$$
which is in the perfect agreement with experimental data \cite{JAGER}.

%!!! Add note about asymptotic at $q^2->0$ and the sum of $\alpha+\beta$ !!!}
%The relatively large $\chi^2$ is due to very small experimental errors and it even becomes slightly larger reaching $2.3$ if only the modern SND data are taken into account. On the Fig.\ref{fig:4} we can see two resonances fit perfectly matches with data (solid line).% From the table~\ref{tbl:1} one can see that even small deviation from $\alpha=\beta=0.5$ lead to the significantly worse description of the experimental data.
%This result is consistent with the VMD model, where photon is saturated by $\rho$ meson and higher vector mesons($\omega$ etc.).
%Based on the result of the 2 resonances fit, must be, in order to describe peak in our approach, one should obtain two close standing poles in the real part of time-like pion TFF. In the imaginary part widths of $\rho$ and $\omega$ need to be finite.

\subsection{$\phi$ peak}\label{peak2}

\begin{figure}[h!]
\includegraphics[width=1.0\textwidth]{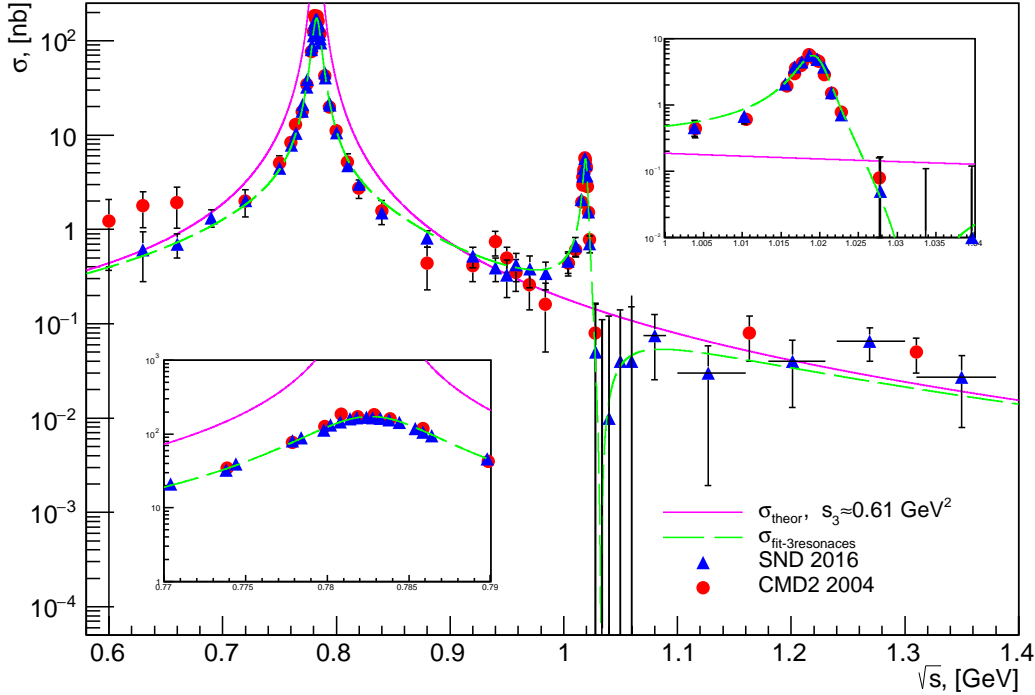}
\caption{Fit of the whole data range. The inserts are zoom in $\rho-\omega$ and $\phi$ peaks.}
\label{fig:5}
\end{figure}

Let us now consider the second peak. Theoretical estimations shows(
work is now in progress), that the such kind of a peak can appear,
if one takes into account $\pi-\eta-\eta'$ mixing and s-quark masses
into the anomaly sum rules. Thus, in this case the pion TFF will
have 3 terms. So, in addition to the terms corresponding to $\rho$ and $\omega$ mesons, one should add a term corresponding to the $\phi$ meson ( the value of $m_{\phi}=1.0194 \ GeV$ and the value of $\Gamma_{\phi}=0.00426 \ GeV$ were taken from PDG \cite{PDGphi}. ). Let us suppose, that $s_{3\phi}$ will be close to the $m_{\phi}^2$ and $m_{\rho}=m_{\omega},s_{3\rho}=s_{3\omega}=m_{\rho}^2=m_{\omega}^2=s_3\approx0.61 \ GeV^2.$ Thus, we got:
\begin{equation}
\label{3resff}
F(q^2)=\frac{1}{2\sqrt{2}\pi^2f_{\pi^0}}\left(\alpha\frac{s_{3}}{s_{3}-q^2+im_{\rho}\Gamma_{\rho}}+\beta\frac{s_{3}}{s_{3}-q^2+im_{\omega}\Gamma_{\omega}}+\gamma\frac{s_{3\phi}}{s_{3\phi}-q^2+im_{\phi}\Gamma_{\phi}}\right).
\end{equation}
Substituting the eq.\eqref{3resff} to \eqref{asr_cs}, one obtains the equation for the total cross section. 
%The equation for the total cross section is obtained:
%\begin{eqnarray}\label{3rescs}
%\sigma_{fit-3resonances} & = & \frac{\alpha_{QED}^3}{12\pi^{2}f_{\pi^0}^2}\Big(\alpha^2\frac{s^2_{3\rho}}{(s_{3\rho}-q^2)^2+m_{\rho}^2\Gamma_{\rho}^2}+\beta^2\frac{s_{3\omega}^2}{(s_{3\omega}-q^2)^2+m_{\omega}^2\Gamma_{\omega}^2}+\gamma^2\frac{s^2_{3\phi}}{(s_{3\phi}-q^2)^2+m_{\phi}^2\Gamma_{\rho}^2} + {}\nonumber \\
%             &   & +2\alpha\beta s_{3\rho}s_{3\omega}\frac{(s_{3\rho}-q^2)(s_{3\omega}-q^2)+m_{\rho}\Gamma_{\rho}m_{\omega}\Gamma_{\omega}}{((s_{3\rho}-q^2)^2+m_{\rho}^2\Gamma_{\rho}^2)((s_{3\omega}-q^2)^2+m_{\omega}^2\Gamma_{\omega}^2)}+ {}\nonumber \\
%             &   & +2\alpha\gamma s_{3\rho}s_{3\phi}\frac{(s_{3\rho}-q^2)(s_{3\phi}-q^2)+m_{\rho}\Gamma_{\rho}m_{\phi}\Gamma_{\phi}}{((s_{3\rho}-q^2)^2+m_{\rho}^2\Gamma_{\rho}^2)((s_{3\phi}-q^2)^2+m_{\phi}^2\Gamma_{\phi}^2)}+ {}\nonumber \\
%             &   &+2\beta\gamma s_{3\omega}s_{3\phi}\frac{(s_{3\omega}-q^2)(s_{3\phi}-q^2)+m_{\omega}\Gamma_{\omega}m_{\phi}\Gamma_{\phi}}{((s_{3\omega}-q^2)^2+m_{\omega}^2\Gamma_{\omega}^2)((s_{3\phi}-q^2)^2+m_{\phi}^2\Gamma_{\phi}^2)}\Big).
%\end{eqnarray}
The result of the fit is shown on the Fig.\ref{fig:5}, values of the fit coefficients are: $\alpha  = 0.556, \ \beta = 0.49, \ \gamma=-0.036$. The $\chi^2$ values are in the Table\ref{tbl3}. 

\begin{table}[h!]
\caption{Values of $\chi^2$ for fit with 3 resonances.}
\label{tbl3}
\begin{center}
\begin{tabular}{|c|c|c|c|}
\hline
 & \textbf{CMD2} & \textbf{SND2016} & \textbf{CMD2+SND2016} \\
\hline
$\chi^2/d.o.f.$ & 2.53 & 1.52 & 1.87 \\
\hline
\end{tabular}
\end{center}
\end{table}
As one can see from the Fig.\ref{fig:5}, the fit with 3 resonances gives good description of experimental data. In the same way, as was discussed in the end of the previous section, by use of \eqref{3resff} we can obtain in the limit $q^2\rightarrow0$  the value 
$$\Gamma(\pi^0\rightarrow2\gamma)\approx7.9 \ eV$$ 
in good agreement with experiment \cite{JAGER}. We found that the coefficient $\gamma$ is negative, which is dictated by the interference term in the vicinity of $\phi$ peak. Moreover, allowing for the phase shift between $\phi$ and $\rho,\omega$ to be close to $\pi$, supporting negative $\gamma$.%the respective angle appears

Let us pay attention to the coefficient $\gamma$, it is found to be
small. As it was mentioned before in the text, the third term in
\eqref{3resff} can appear due to effects of $\pi-\eta-\eta'$ mixing.
So the coefficient $\gamma$ corresponds to the mixing angle
$\theta_{\pi-\eta}$. Theoretical prediction of the mixing angle
value $\theta_{\pi-\eta}$ was done in the works  \cite{IOFFE}, \cite{GTW} (and also see \cite{IO}:
\begin{equation}\label{angle}
\theta_{\pi-\eta}
=\frac{1}{\sqrt{3}}\frac{m_u-m_d}{m_u+m_d}\cdot\frac{m^2_{\pi}}{m^2_{\eta}}
=- 0.0150 \pm0.020.
\end{equation}
Note, that the values of the coefficient $\gamma$ and the
mixing angle $\theta_{\pi-\eta}$ are consistent by the order of
magnitude. This consistency provides additional support to our
assumption about origin of the third term of \eqref{3resff}.

Due to the fact that $\phi$ meson has very narrow width, the
contribution of the third term is non-negligible only at the region
of the $\phi$ resonance, and thus, it can be safely neglected everywhere, except resonance region. Note, that the fit of total cross section with 3 resonances(dashed line) differs from the total cross section with a single pole \eqref{asr_cs}(solid line) by $\approx10\%$ at $\sqrt{s}=1.3 \ GeV$. And the difference between them decreases with the increasing of the $q^2$, thus, as we mentioned earlier in the text, in region far from the poles the total cross section with resonances coincides with the single pole total cross section.  
 %(See Fig.\ref{fig:6} in appendix. Dashed curve corresponds to the eq.\eqref{3rescs} with $\gamma=0$, solid curve correspond to the eq.\eqref{asr_cs}).

As the result of the analysis of the fits, one can conclude that the
modified equation for the pion TFF \eqref{3resff} and the
corresponding total cross section shows rather good
agreement with the experimental data. The whole spectrum of the data
probably can be described if one takes into account the mixing in
axial channel. Thus, in the modified ASR approach, which will be
taking into account the effects of the $\pi-\eta-\eta'$ mixing and
small corrections (as well pertubative as non-pertubative), the
equation of the pion TFF will contain three terms. Theoretical estimations show that  the coefficients $\alpha,\beta$ and $\gamma$ of the eq.\eqref{3resff} can be expressed in terms of the mixing angles $\theta_{\eta-\eta'}$, $\theta_{\pi-\eta}$, $\theta_{\pi-\eta'}$. The results of the fits provide hard restrictions to the expected values of $\pi-\eta-\eta'$ mixing angles and parameters of the modified ASR approach. The work now is in progress.

\section{Conclusions and Outlook}\label{last}

1. We show that the result obtained in the paper \cite{KOT2013} for
the pion timelike TFF \eqref{ff}, can be used to describe the data in the
regions far from the pole, and the place of the pole coincides with
the experimental peak. 
It has been shown, that eq.\eqref{ff}, and the corresponding equation of the total cross section eq.\eqref{asr_cs_theor}, has better agreement with the data if $s_3\approx0.61 GeV^2$.
This one coincides with the result of matching ASR and LCSR \cite{OPST} that $s_3$ should vary between $0.67 \ GeV^2$ at
large $q^2$ and $0.61 \ GeV^2$ at low $q^2$.

2. We propose a modification of the equation of the pion TFF in
order to describe the data at the resonance regions. As a result of
the analysis, we may conclude that in order to describe the whole
spectrum of data, one should obtain formula of the pion TFF
containing three terms. Using modified equation of  the pion TFF
\ref{3resff}, we perform fits of the experimental data and obtain
the values of the fit coefficients. The obtained result provides a good description of the experimental data, including the correct limit for the case of real photons($q^2\rightarrow0$).

3. To obtain three terms in the pion TFF within the ASR approach one should include the effects of the $\pi-\eta-\eta'$ mixing. Let us stress, that the obtained value of the fit coefficient $\gamma$  corresponding to the third term \eqref{3resff}  has the same order of the magnitude as the $\theta_{\pi-\eta}$ mixing angle.
So it confirms our assumptions. 
The achieved values of the fits coefficients lead to strong restrictions
to the values of the $\pi-\eta-\eta'$ mixing angles and can be used
in matching with the theoretical values, calculated, in particular, in the
modified ASR approach.

The work was supported in part by RFBR grant 14-01-00647.

\begin{comment}
%\newpage
%\section{Appendix}
%\begin{center}
%\begin{table}[h!]
%\caption{Dependence of the $\chi^2(\alpha,\beta)$ function from the values of the $\alpha$ and $\beta$.\label{tbl:1}}
%\begin{ruledtabular}
%\begin{tabular}{|c|c|c|}
%\hline
%$\alpha$ & $\beta$ & $\chi^2/d.o.f.$ \\
%\hline
%1 & 0 & 745.32 \\
%\hline
%0.95 & 0.05 & 706.53 \\
%\hline
0.9 & 0.1 & 646.54 \\
\hline
0.85 & 0.15 & 568.37 \\
\hline
0.8 & 0.2 & 476.20  \\
\hline
0.75 & 0.25 & 375.31 \\
\hline
0.7 & 0.3 & 272.12 \\
\hline
0.65 & 0.35 & 174.21 \\
\hline
0.6 & 0.4 & 90.28 \\
\hline
0.55 & 0.45 & 30.17 \\
\hline
0.5 & 0.5 & 4.84 \\
\hline
0.45 & 0.55 & 26.40 \\
\hline
0.4 & 0.6 & 108.11 \\
\hline
0.35 & 0.65 & 264.34 \\
\hline
0.3 & 0.7 & 510.60 \\
\hline
0.25 & 0.75 & 863.55 \\
\hline
0.2 & 0.8 & 1340.99 \\
\hline
0.15 & 0.85 & 1961.82 \\
\hline
0.1 & 0.9 & 2746.11 \\
\hline
0.05 & 0.95 & 3715.06 \\
\hline
0 & 1 & 4891 \\
\hline
\end{tabular}
\end{table}
%\end{ruledtabular}
\end{center}

\begin{figure}
\includegraphics[width=1.0\textwidth]{gamma=0.eps}
\caption{Fit of the whole data with $\gamma=0$ in the
eq.\eqref{3rescs}} \label{fig:6}
\end{figure}
\end{comment}

\end{document}